\documentclass[12pt,preprint]{aastex}
\newcommand{\mpc}{h^{-1}Mpc}
\newcommand{\gsim}{\mathrel{\hbox{\rlap{\hbox{\lower4pt\hbox{$\sim$}}}\hbox{$>$}}}}

\shorttitle{2dF and SDSS Galaxy Density Profiles}
\shortauthors{D\'{\i}az et al.}

\begin{document}

\title{2dFGRS and SDSS Galaxy Group Density Profiles}

\author{Eugenia D\'{\i}az\altaffilmark{1,2}, Ariel Zandivarez\altaffilmark{2}, Manuel E. Merch\'an\altaffilmark{2} and Hern\'an Muriel\altaffilmark{2}}
\affil{Grupo de Investigaciones en Astronom\'{\i}a Te\'orica y Experimental,
IATE, Observatorio Astron\'omico, Laprida 854, C\'ordoba, Argentina.}

\altaffiltext{1}{Agencia Nacional de Promoci\'on Cient\'{\i}fica}

\altaffiltext{2}{Consejo de Investigaciones Cient\'{\i}ficas y T\'ecnicas de la Rep\'ublica Argentina.}

\begin{abstract}
We have analysed the distribution of galaxies in groups identified in the
largest redshift surveys at the present: the final release of the 2dF Galaxy 
Redshift Survey and the first release of the Sloan Digital Sky Survey. 
Our work comprises the study of the galaxy density profiles and the fraction
of galaxies per spectral type as a function of the group-centric distance.
We have calculated the projected galaxy density profiles of galaxy groups 
using composite samples in order to increase the statistical significance of 
the results. Special cares have been
taken in order to avoid possible biases in the group identification and the
construction of the projected galaxy density profile estimator.  
The results show that the projected galaxy density profiles obtained for
both redshift surveys are in agreement with
a projected Navarro, Frenk \& White predictions in the range $0.15< r/r_{200}
 < 1$, whereas a good fit for the measured profiles in the whole range of 
$r/r_{200}$ is given by a projected King profile. 
We have adopted a generalized King profile to fit the measured projected 
density profiles per spectral type. In order to infer the 3-D galaxy density 
profiles, we deproject the 2-D density profiles using a deprojection  
method similar to the developed by Allen \& Fabian. From 2-D and 3-D galaxy 
density profiles we have estimated the corresponding galaxy fractions per 
spectral type. The 2-D fraction of galaxies computed using the projected 
profiles show a similar segregation of galaxy spectral types 
as the obtained by Dom\'{\i}nguez et al. for groups in the 
early data release of the 2dF Galaxy Redshift Survey. 
As expected, the trends obtained for the 3-D galaxy fractions show 
steeper slopes than the observed in the 2-D fractions.
\end{abstract}


\section{Introduction}
Several models have been proposed to characterize the projected
galaxy density in clusters of galaxies. Most of these models assume
spherical symmetry and that the matter distribution is traced by galaxies. The first assumption
can be true for a sub-sample of clusters, while the second is more
difficult to quantify and is close related with different processes like
galaxy formation, galaxy evolution, dynamical friction, etc.  Assuming
that the galaxy velocity dispersion is well represented by an isothermal sphere,
King (1962) proposed a model to describe the galaxy projected density profile.
More recently, Navarro, Frenk and White (1995, hereafter NFW95) analysing high-resolution
N-body simulations propose an universal
profile for dark matter halos.  These authors found that their model can
appropriately describe the mass density profiles for a large range of masses.
The observational evidence coming from the giant arcs in clusters can be used
to introduce constrains to the  mass distribution in the core  of clusters 
(Navarro, Frenk and White 1997). Strong lensing effects require a very small 
core radii, that in principle can be  consistent with the NFW mass profile. 
Nevertheless, very high resolution cosmological simulations  produce density
profiles with inner slopes $\sim -1.5$ that are steeper than the obtained from 
NFW (slope of $\sim -1$ near the center) (Moore et al. 1998).
It is not clear whether  the galaxy density profile will
follow the mass, in particular in the very core of  clusters where the
scales of galaxies impose the resolution limit. The controversy
among different models to describe both, the mass and galaxy profile is 
still open and more observational evidences are needed.

Adami et al. (1998) studied the galaxy density profiles for an important
sample of rich clusters of galaxies.  One of the main clue of this work was to
investigate whether the galaxy distributions have cores (like King profile) or
cups (NFW profile). Based on redshift information taken from ENACS (ESO Nearby
Abell Cluster Survey, Katgert et al. 1998) and projected galaxy distribution 
coming from COSMOS (MacGillivray \& Yentis, 1994) Adami et al. 1998 conclude 
that  in general the King profile provides a better representation of the data 
than the NFW profile.

Bartelman (1996) derived  the analytic expression for the 
surface mass density of the NFW profile while Lokas \& Mamon (2001) provide 
the tools for modeling the NFW profile and give predictions for different 
observational quantities.
Independently of any analytical model the  3D profile can be derived applying a
deprojection method similar to those implemented for the X-ray analysis of 
the hot intra-cluster gas (Allen \& Fabian 1997).

Most of the previous analysis on galaxy density profiles consider galaxies 
regardless their properties. The effect of morphological segregation
(MS), (Dressler 1980, Whitmore \& Gilmore 1991, Dominguez, 
Muriel \& Lambas 2001)
implies that different galaxy populations will have different galaxy density 
profiles. MS works were carried out analysing the bidimensional (2-D) galaxy 
fractions of different morphological types.
In order to recover the 3D MS, the 
spatial (3-D) galaxy density profiles for each morphological type should 
be previously known. 
Salvador-Sole and Sanrom\`a (1989) have analysed the observed 
correlation between morphological fractions and the projected density of 
galaxies. They found that this correlation is a consequence of an intrinsic 
3D effect that is dependent on cluster concentration.

Using the  Merch\'an \& Zandivarez (2002) group catalog constructed from
the early data release of the 2dF Galaxy Redshift Survey, Dominguez et al. 
(2002) derive the relative fraction of galaxies with different spectral types
as a function of local galaxy density and group-centric distance. These 
authors found that for high mass groups ($M_V \gsim 10^{13.5} M_\odot$) 
a strong dependence of the relative fraction of spectral types on 
both, galaxy density and group-centric radius is observed. 

The aim  of this paper is to determine the projected and the 3-D galaxy group
density profiles for different spectral types. 
We also derive the intrinsic 3-D MS that 
results from the observed 2-D MS. The observational results are compared with 
those obtained from different analytical models.
This paper is outlined as follows: the deprojection method to apply on 
projected density profiles is described in section 2. The galaxy and group 
data in the 2dF Galaxy Redshift Survey and the Sloan Digital Sky Survey is 
quoted in section 3. The mock catalog tests made for the projected density 
profile estimator and the subsequent application to observational data
are carried out in section 4. The derivation of the 3-D density 
profiles and the estimation of 3-D galaxy fractions are described in section 5.
Finally, in section 6 we summarize our conclusions.

\section{Density profiles in numerical simulations}
\subsection{Density profiles estimator}
We use collisionless cosmological numerical simulations of flat, low density, 
cold dark matter universes performed using the Hydra N-body code 
developed by Couchman et al. (1995). Simulations are constructed with $128^3$
particles in a cubic comoving volume of $180 \ h^{-1} \ Mpc$ per side starting 
at $z=50$. The adopted cosmological model corresponds to an universe with a 
present day matter density $\Omega_m=0.3$, vacuum energy density 
$\Omega_{\Lambda}=0.7$, initial spectral slope $n=1$, 
$\Gamma=0.21$, Hubble constant $H_0=100 \ h \ km \ s^{-1} \ Mpc^{-1}$ with 
$h=0.7$ and an amplitude of mass fluctuations of $\sigma_8=0.9$.

Groups used in this section were, initially, identified using a standard 
friend of friend finder algorithm  with density contrast of $\delta \rho / 
\bar{\rho} =300$ corresponding to a linking length of $d_0=0.15\times n^{-1/3}$,where $n$ 
is the mean density number of particles; after that we select high mass groups, 
spanning a mass range from $3.5\times10^{13} $ to $5.8\times10^{14} \ M_\odot \ 
h^{-1}$. The final sample consists of $690$ groups.

In an attempt to increase the statistical significance, we combine all groups 
to produce a composite set of dark matter (DM) particles, 
properly scaled to take into account the different group sizes and masses. 
The composite sample was made assuming that all groups obey the same 
type of density profile but with different scales. Hence, it is necessary
to introduce two parameters: one of them to normalize the group-centric distances and the other to normalize the masses.  
For convenience, we adopt the radius at which the mean interior density 
is 200 times the mean density of the universe ($r_{200}$) as the normalization scale and the mass contained in $r_{200}$ ($M_{200}$) as the 
mass normalization.
 
We measure the projected density profile for the composite sample 
as a function of the normalized radii $r/r_{200}$. The binning scheme used 
through all this work is the equal number binning. The measured density profile
will be compared with the analytical function obtained by NFW97.
The 3-D NFW97 density profile is described by the following equation:

\begin{equation}
\frac{\rho (r)}{\rho_c}=\frac{\delta_c}{c\frac{r}{r_{200}}(c\frac{r}{r_{200}}+1)^2}
\label{NFW}
\end{equation}
where  
$\rho_c=3 H^2 / 8 \pi G$ is the critical density for closure, $c$ is the 
concentration of the halo, and $\delta_c$ is the characteristic density 
(see eq. 2 of NFW97). Through all this work we use 
a  projection of the equation \ref{NFW} obtained by numerical integration 
along the line of sight.
 
\begin{figure}
\epsscale{0.80}
\plotone{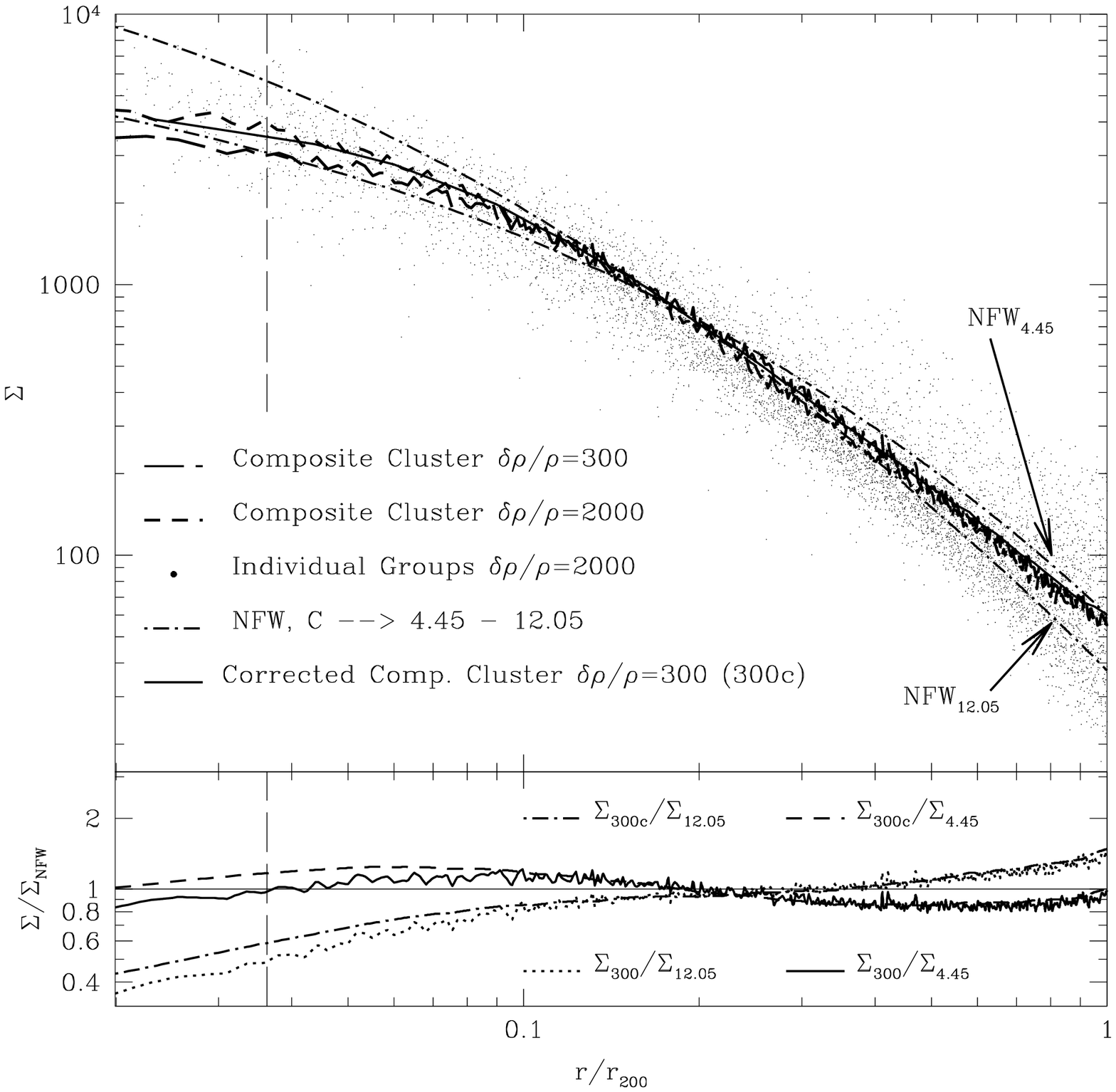}
\caption{
Upper panel: Projected density profiles as a function of the 
normalized group-centric 
distance. Long-dashed line is the measured projected density profile for the
composite sample using groups identified with $\delta \rho/\bar{\rho}=300$ 
while short-dashed line is the corresponding for groups obtained using 
$\delta \rho/\bar{\rho}=2000$. The solid line is the measured projected density 
profile for groups identified with $\delta \rho/\bar{\rho}=300$ after a correction
made on the group center location. Single group projected density profiles 
for $\delta \rho/\bar{\rho}=2000$ are shown as points. Projected NFW profiles 
are drawn with dot-dashed lines using $c$ of $4.45$ and $12.05$.
Lower panel: Ratios between measured profiles and 
 NFW predictions as a function of $r/r_{200}$. The key of each curve is 
included in the figure. The subscripts represent the $\delta \rho/\bar{\rho}$ 
and the $c$ parameter for the measured profiles and the NFW respectively.  
\label{fig1}
} 
\end{figure}

Upper panel of Figure \ref{fig1} shows the projected density profile 
normalized to the number 
of groups involved in each bin (long dashed line) measured for the composite 
sample. The dot-dashed lines are the projected NFW profiles corresponding to 
different values of the $c$ parameter ($4.45$ and $12.05$). 
These $c$ values are associated with a wide range of masses 
($ 10^{11}<M/M_\odot<10^{15}$).
In the lower panel it can be seen, in long dashed line, the comparison 
between the measured profile and the analytical NFW profiles, plotted as the 
ratio $\Sigma/\Sigma_{NFW}$. 
Left to the vertical line in Figure \ref{fig1} the densities are underestimated 
due to the uncertainty in the location of the group geometric center.
An improvement of the group geometric center estimation can be obtained 
increasing $\delta \rho/\bar{\rho}$ in the group 
identification, which produces groups with geometric centers closer to the 
overdensity peaks. The measured projected density profile for groups identified 
with $\delta \rho/\bar{\rho}=2000$ is plotted in the upper panel of Figure 
\ref{fig1} (short-dashed line). 
This profile has a very good agreement with the NFW predictions. 
We also show in dots the density profiles for each group.
This procedure for improving the group geometric center is not feasible in 
observational surveys, since the number of groups identified is strongly 
decreased for high overdensities and this affects the reliability of the 
results. Hence, it is important 
to find another method to correct the group center positions and 
keeping constant the number of groups. 
The procedure adopted for the estimation of the new group centers
takes into account the projected local number densities at the position of each 
particle (galaxy, when dealing with observational data). A first estimation 
of the center is obtained by the following equation:
\begin{equation}
r_c^{(1)}=\frac{\sum_{j=1}^N \rho_j^{PL}r_j}{N \bar{\rho}^{PL}}
\label{rcentro}
\end{equation}
where $N$ is the number of particle members of each group, 
$\rho_j^{PL}$ is the projected local density in the position of the $j^{th}$ 
particle and $\bar{\rho}^{PL}$ is the mean projected local density. 
The projected local density for the $j^{th}$ particle is computed 
using the projected circular area which contains the $n$ nearest particles. 
We use the values of $n=75$ in the simulation and $n=5$ in the 
catalogs (Dom\'{\i}nguez et al. 2002). 
This procedure improves the center location, 
but in some cases it is not enough to match the identified group centers with 
the corresponding overdensity peaks.
Consequently, we adopt an iterative procedure as follow:
\begin{enumerate}
\item Using the geometric center position ($r_b$) we determine the distance $d_0$ to 
the farthest particle/galaxy.
\item After the computation of $r_c^{(1)}$ we reject all the 
particles/galaxies with distances to $r_c^{(1)}$ greater than $d_0$. Then we  
estimate $d_1$ for the remaining particles/galaxies.
\item We calculate $r_c^{(2)}$ for the new group using the equation 
\ref{rcentro} and applying the same procedures as described in item 2. 
\end{enumerate}  
 
The iteration must go on until 
$d_{M-1}=d_{M}$. Finally, after $M$ iterations, the group center obtained is $r_c=r_c^{(M)}$.
Besides the improvement in the determination of the center position, 
the proposed method also correct the group 
merging problem produced by the process of identification, 
in other words, groups with two or more 
overdensity peaks are cleaned, preserving the highest peak.

We use this method to determine the centers for the groups identified with $\delta \rho/\bar{\rho}=300$. The density profile obtained for the corrected 
sample is 
shown in solid line in the upper panel of Figure \ref{fig1}. 
It can be seen that our center 
estimator allows us to reproduce the density profile obtained using 
$\delta \rho/\bar{\rho}=2000$, with the advantage of keeping constant 
the number of groups identified with $\delta \rho/\bar{\rho}=300$. 
The agreement with the NFW predictions 
can also be observed in the lower panel of this Figure (solid line curves).

\subsection{Deprojection method of density profiles}
\begin{figure}
\plotone{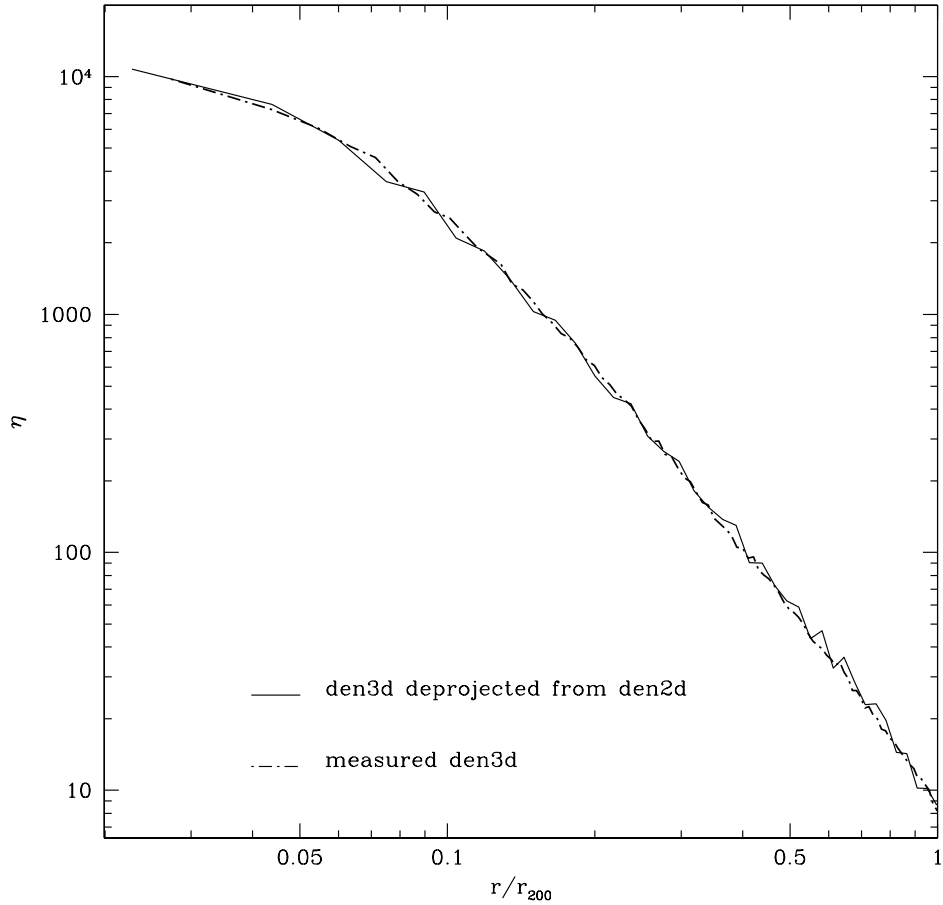}
\caption{
Deprojected density profiles for DM groups. The 3-D density profile calculated 
from the 2-D density profile using the deprojection method is shown as solid 
line. Dot-dashed line corresponds to 3-D density profile measured directly 
for the composite sample.
\label{fig2}
} 
\end{figure}

From the projected density profile we calculate the 3-dimensional profile 
applying a deprojection method similar to that developed by Allen \& Fabian
(1997). The deprojection analysis assumes spherical symmetry. 
The method is a matter of dividing the spatial 
distribution into a series of $n$ concentric spherical shells. 
The projected number of galaxies $N(j)$ in the $jth$ cylindrical bin 
can be considered as the contribution of different spherical shells, which 
can be calculated as 
the 3-D numerical density of each shell, $\eta(j)$, multiplied by 
the corresponding volume $V_{j,i}$:
\begin{eqnarray}
N(j)=\sum^{n}_{i=j}\eta(i)V_{j,i} &; \ 1<j<n \ , &  j<i<n
\label{denreal}
\end{eqnarray}
where $V_{j,i}$ is the volume corresponding to the intersection 
of a spherical shell with inner radius 
$r_{i-1}$ and outer radius $r_{i}$ and a cylindrical shell with radii $r_{j-1}$
(inner) , $r_{j}$ (outer). 
Since the projected density in the last cylindrical shell is only dependent on 
the 3-D density of the outer spherical shell, equation \ref{denreal} can be 
used to obtain the 3-D density profile $\eta (j)$ from outer to inner shells:
\begin{equation}
\eta(j)=(N(j)-\sum^{n}_{i=j+1}\eta(i)V_{j,i})/V_{j,j}
\label{denreal3d}
\end{equation}
In order to test the method reliability we apply it to the composite sample 
made with DM groups identified in N-body simulations. 
The derived 3-D density profile is shown in Figure 
\ref{fig2} (solid line) and it is compared with the 3-D profile 
directly measured in simulations. From this comparison we can observe a perfect
recover of the 3-D density profile. 

\section{The data}
\subsection{The galaxy sample}
At present, the largest samples of galaxies with spectroscopic redshift 
determinations are the 2dFGRS (2 degree Field Galaxy Redshift Survey) and the 
SDSS (Sloan Digital Sky Survey). In this work we use both catalogs in order 
to obtain the largest samples of groups and increase the reliability of our 
results.
 
The 2dF survey covers 1500 $deg^2$ with a median depth of $\bar{z}=0.11$. The 
complete 2dFGRS consists of $221414$ galaxies in two declination strips and 
2-degree random fields scattered around the southern galactic pole (SGP) strip.
The galaxies were taken from an improved version of the APM galaxy survey 
(Maddox et al., 1990a,b; Maddox, Efstathiou \& Sutherland, 1996).
The sky coverage of the 2dFGRS is not uniform (a detailed completeness 
description is given by Colless et al., 2001; see also http://www.mso.anu.edu.
au/2dFGRS/). 
Galaxies in this survey also have a spectral classification given by the 
parameter $\eta$ based on a principal component analysis as described by 
Madgwick et al. (2002).  

Recently, the Sloan Digital Sky Survey has validated and made publicly 
available the First Data Release (Abazajian et al. 2003) which is a photometric
and spectroscopic survey constructed with a dedicated $2.5 \ m$ telescope at
Apache Point Observatory in New Mexico. The First Data Release consist of
$2099 \ deg^2$ of five-band ($u \ g \ r \ i \ z$) imaging data and 186240 
spectra of galaxies, quasars and stars. In this work we mainly use the 
spectroscopic sample. The SDSS Team has found that 
the survey redshift accuracy is better than $30 \ km \ s^{-1}$.
Our sample comprises 100118 galaxies with radial velocities spanning the range
$420 \ km \ s^{-1} \leq V \leq 90000 \ km \ s^{-1}$ and an upper apparent
magnitude limit of 17.77 in the r-band.
In order to work with different kinds of galaxy population, we compute
a galaxy spectral type based on a Principal Component Analysis (PCA),
using a cross-correlation with eigentemplates constructed from SDSS
spectroscopic data. These spectral types are computed with the first two
eigencoefficients as recommended by de SDSS Team.  

\subsection{The group samples}
The group samples obtained from the 2dFGRS and SDSS were constructed using
an algorithm similar to that developed by Huchra \& Geller (1982).
Particularly, we have introduced some modifications in the group finder
in order to take into account the sky coverage of the 2dFGRS. The adopted
procedure is the same as described by Merch\'an \& Zandivarez (2002) who
consider the redshift completeness, magnitude limit and $\mu$ masks of the
2dFGRS. The identifications were carried out using a density contrast of
$(\delta \rho /\bar{\rho})_z=80$ and a line of sight linking length of  
$V_0=200 km \ s^{-1}$. 

\begin{figure}
\plotone{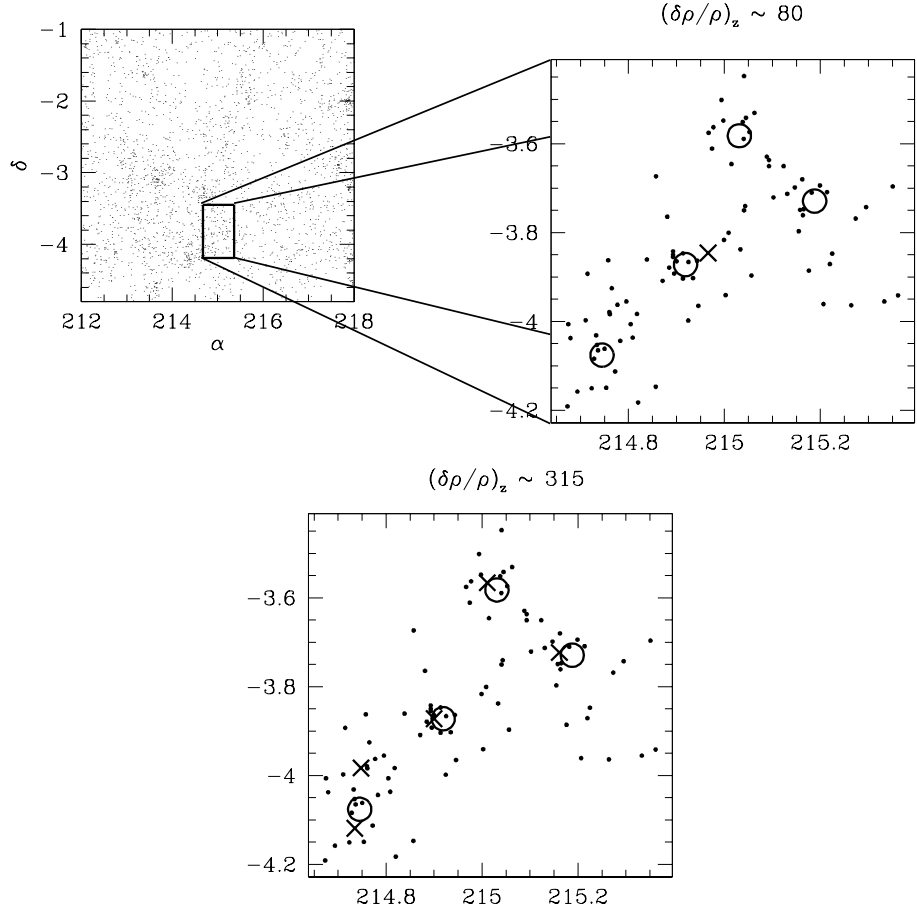}
\caption{Comparison of groups identified in real and redshift space.
Right upper panel shows, for a given patch on the sky, groups identified in 
real space with open circles while groups identified in redshift space are 
drawn with crosses. These identifications were performed using the same
density contrast ($\delta \rho/\bar{\rho} \sim 80$). Dots are the galaxies
belonging to the groups identified in redshift space.
Lower panel shows a similar comparison as showed in the previous panel but now
crosses denote the resulting sample of groups in redshift space after 
performing a second identification on the first sample of groups using 
$\delta \rho/\bar{\rho} \sim 315$.  
\label{fig3}
} 
\end{figure}

As was detailed in section 2, the group center location has an important 
influence on the density profiles estimations. It is known that working 
on observational redshift surveys means that group identification must be 
performed in redshift space. This sort of procedure could induce 
missidentifications of groups respecting to those that would be identified
in real space. For instance, the group finder algorithm in redshift space 
 can not completely eliminate the interloper
effect on the identification. This effect is likely to produce an artificial
increment in the projected size of groups or the detection of fictitious
systems with multiple overdensities. 
In order to understand the relation between groups identified in real and redshift space we perform a comparative study using mock catalogs 
(see section 4 for a detailed description of mock catalogs construction). 
The groups identification in real space was performed 
using a similar algorithm as the adopted for redshift space, but using 
the same linking length in both, transverse and radial directions.
Right upper panel in Figure \ref{fig3} shows a comparison among the groups 
identified in both, real space (open circles) and redshift space (crosses) using
$\delta \rho/\bar{\rho}=80$, for a given patch in the sky. Points represent the 
galaxies/particles in this region identified as group members in redshift space. As can 
be seen, several groups in real space were joined in a single group in 
redshift space. 
Our purpose is to reproduce the groups identified 
in real space with $(\delta \rho/\bar{\rho})_r=80$. 
In order to do this, we have carried out a second identification 
on the previous group sample identified in redshift space, 
varying the density contrast, $(\delta \rho /\bar{ \rho})_z$, 
until we observe a similar identification as the obtained in real space.
Lower panel in Figure \ref{fig3} shows the same comparison as the one plotted 
in the right upper panel, but here crosses are the groups obtained 
after a second identification in redshift space with a density contrast
of $(\delta \rho /\bar{\rho})_z\sim315$.
Even though the second identification does not perfectly reproduce the one 
obtained in real space, our study indicates that the $(\delta \rho /\bar{\rho})_z$ 
adopted for the second identification is the best choice to produce a sample 
of groups quite similar to the observed in real space.
Consequently, we have performed a second identification, using the best 
density contrast value previously obtained, over the group samples of the 
2dFGRS and SDSS described before. Finally, the group centers were computed 
using the iterative method detailed in section 2. 
The group samples used through this work include systems with masses 
greater than $6 \times 10^{13} h^{-1} M_{\odot}$ and 
having more than 10 galaxy members. The adopted mass threshold, only selects 
the more massive groups that are the most interesting when spectral type 
properties are studied (Dom\'{\i}nguez et al., 2002). The final samples 
comprise $132$ groups for the 2dF and $86$ for the SDSS.
The group physical properties were computed using the same formulas adopted 
by Merch\'an \& Zandivarez (2002). The median group properties and the widths 
of the distributions (semi-interquartile range) are quoted in Table 
\ref{proptab}. Analysing the information shown in this table it can be seen 
that the average properties are very similar for both catalogs. This is an 
expected result taking into account the similarities of both catalogs. 

\section{Projected galaxy density profiles}
In order to measure projected galaxy density profiles we use a similar 
procedure to that employed in the simulations taking into account the surveys 
limitations. We construct the composite samples for both group catalogs 
adopting $r_{200}$ as the normalization scale and $M_{200}$ as masses 
normalization. The computation of $r_{200}$ was made following Carlberg et 
al.(1997):
\begin{equation}
r_{200}=\frac{\sqrt{3}}{10}\frac{\sigma}{H(z)}
\end{equation}
while $M_{200}$ was obtained using (see appendix of NFW97):
\begin{equation}
M_{200}=\left(\frac{r_{200}}{K}\left( \frac{\Omega_0}{\Omega(z)}\right)^3 (1+z)\right)^3 \ h^{-1} \ M_\odot
\end{equation}
where $K=1.63\times 10^{-5}\mpc$.
These scaling relations are in very good agreement with the properties 
directly measured from the individual DM groups.  

Working with observational samples require to have particular considerations 
when constructing density profiles, specially when seeking for the 
largest statistical sample.
As was noticed by Whitmore, Gilmore \& Jones (1993, hereafter WGJ93), a 
magnitude cutoff decreases the number of galaxies while affects the mix of 
spectral types since the luminosity functions are different for each type.
Distant systems will only include the brightest galaxies 
, which tend to be the earlier, resulting in an incomplete source of 
information.
A magnitude cutoff correction is made following WGJ93 where a weight is given 
to each galaxy. This weight is a function of the redshift, the spectral type and
the catalog apparent magnitude limit, and it is given by the following 
equation: 
\begin{equation}
w(z,m_l)=\left[\frac{\int ^{M_l(z)}_{-\infty}\Phi(M)dM}{\int^{M_l(z_c)}_{-\infty}\Phi(M)dM}\right]^{-1}
\label{w(z)}
\end{equation}
where $\Phi(M)$ is the luminosity function per spectral type of galaxies 
in groups, 
$M_l(z)=m_l-25-5log(d_L)-(k+e)+5log(h)$ the absolute magnitude, $d_L$ is the 
luminosity distance, $m_l$ the catalog apparent magnitude limit and $z_c$ is 
chosen as a typical redshift for groups in the sample.

\subsection{Mock catalogs}

\subsubsection{Angular Masks}
Since the sky coverage of the 2dF group sample is not uniform, we need to make 
extra-corrections before measuring the projected density profiles.
With the aim of determining and testing the corresponding corrections 
to the observed sample, we use four types of mock catalogs. Each of 
these mocks corresponds to different sky coverage as in Merch\'an $\&$ 
Zandivarez (2002). So, we study the influence on the density 
profiles of each distinctive feature present in the catalog. 
To increase the statistical strength we construct a set of ten mock catalogs 
for each type from ten cosmological simulations with different initial 
conditions (section 2). 
Given the periodicity of the simulation box we locate the observer at an
arbitrary position and repeat the box until the survey extent is completed.
These catalogs are constructed using a bias scheme $b=1$ between DM particles 
and galaxies which is quite accurate to reproduce the clustering of the data 
mainly on large scales. Adopting the galaxy luminosity function given by 
Norberg et al. (2002) we assign absolute magnitudes to particles obtaining
mock catalogs with the same selection function than the observed for the 
2dFGRS. 

\begin{figure}
\epsscale{0.80}
\plotone{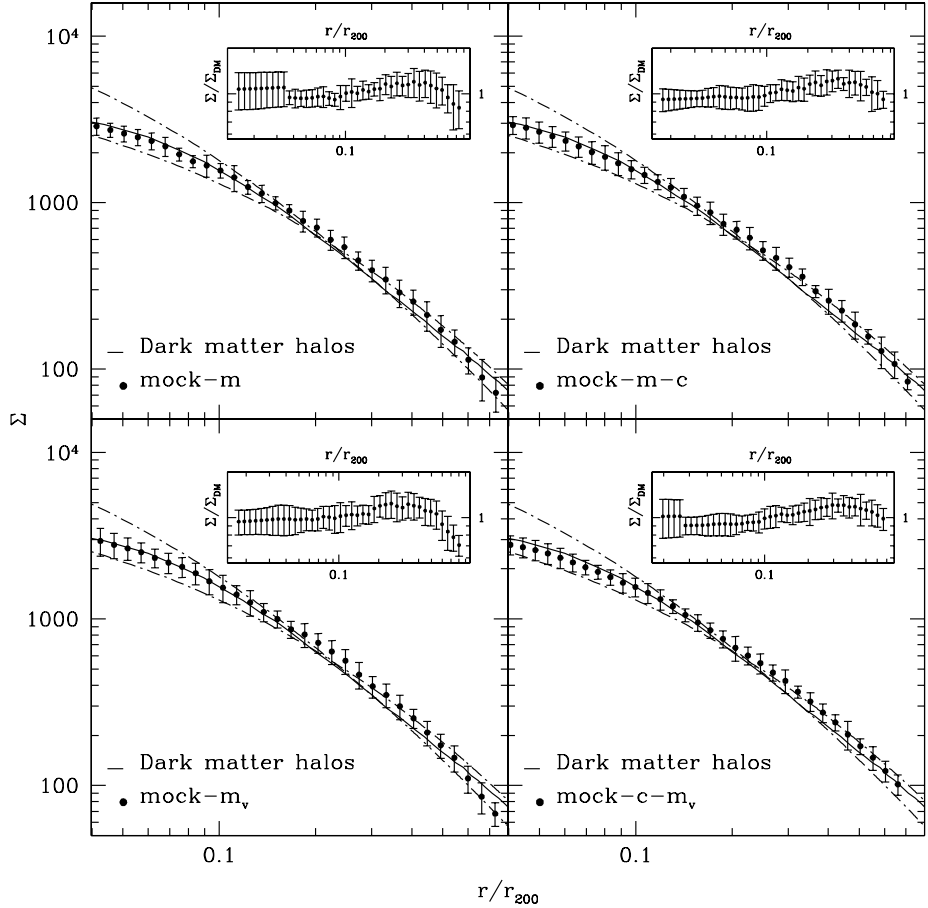}
\caption{
Projected density profiles measured for composite samples from mock catalogs. 
The left upper panel shows with points the averaged projected density profile 
measured for composite samples obtained from 10 mock-m catalogs (see the text).
Right upper, left lower and right lower panels show the same profiles as left 
upper panel but measured for mock-m-c, mock-$m_v$ and mock-c-$m_v$ respectively.
Error bars are computed measuring the dispersion for each set of mock catalogs.
Solid line in each panel corresponds to the projected density profile computed
using groups identified in the dark matter N-body simulations. 
Dot-dashed lines in each panels are projected
NFW density profiles as plotted in Figure \ref{fig1}. The inset panels show
the ratio between the averaged projected density profiles and the projected
density profile obtained from the dark matter halos. 
\label{fig4}
} 
\end{figure}

The first set of mock catalogs (mock-m) introduces a fixed faint survey 
magnitude limit, $m_l=19.2$. We identify groups in this mock in the same way 
as in the 2dF sample (groups with masses greater than $6 \times 10^{13}M_\odot$
, and having more than $10$ members). After constructing the composite sample 
(as explained in section 2) we measure the projected density profile counting
galaxies weighted by the equation \ref{w(z)}. 
The projected density profiles measured for each mock catalog were averaged 
and the corresponding mean profile is shown in the left upper panel in Figure 
\ref{fig4} (points). This profile is compared with the projected density 
profile computed from the DM groups identified in the N-body simulations
(solid line).

The second set of mock catalogs (mock-m-c) has also a fixed faint survey 
magnitude limit but adding the effect of redshift completeness as in the 
real survey. 
The procedure to make the composite sample of groups is the same in all 
cases. At this time, we put another weight to the galaxies in order to measure 
the projected density profile. This weight is the result of multiplying 
$w(z,m_l)$ by the redshift completeness $c(\alpha,\delta)$ available from 
the 2dF mask. In Figure \ref{fig4}, right upper panel shows the 
averaged profile corresponding to this set of mock catalogs (mock-m-c).

The third set of mock catalogs (mock-$m_v$) has a faint survey magnitude 
limit depending on the angular position of a particle consistent with the  
apparent magnitude limit derived from the 2dF mask. 
To measure the projected density 
profile we introduce a weight $w(z,m_v)$ that take into account the variable 
magnitude limit in the calculation of $M_l(z)$.  
The averaged density profile is shown in Figure \ref{fig4} (left lower 
panel).

Finally we use a last set of mock catalogs (mock-c-$m_v$) which has both 
effects, the variable magnitude limit and the redshift completeness 
masks. In this case, the weight assigned to each galaxy consists in the 
multiplication by both weights, $w(z,m_v)$ and $c(\alpha,\delta)$. 
The mean profile is shown in the right lower panel in Figure \ref{fig4}. 
Error bars in Figure \ref{fig4} are computed measuring the 1-$\sigma$
dispersion over each set of ten mock catalogs used to obtain the average 
density profiles. 
The inset panels show the ratio between the averaged projected density
profiles for each set of mock catalogs and the projected density 
profile measured for the DM groups identified in the N-body simulation. 
From these panels we observe that we are able to recover the profiles 
obtained for the simulation making the appropriated corrections on each 
mock catalog.

\subsubsection{Missing Pairs}

\begin{figure}
\plotone{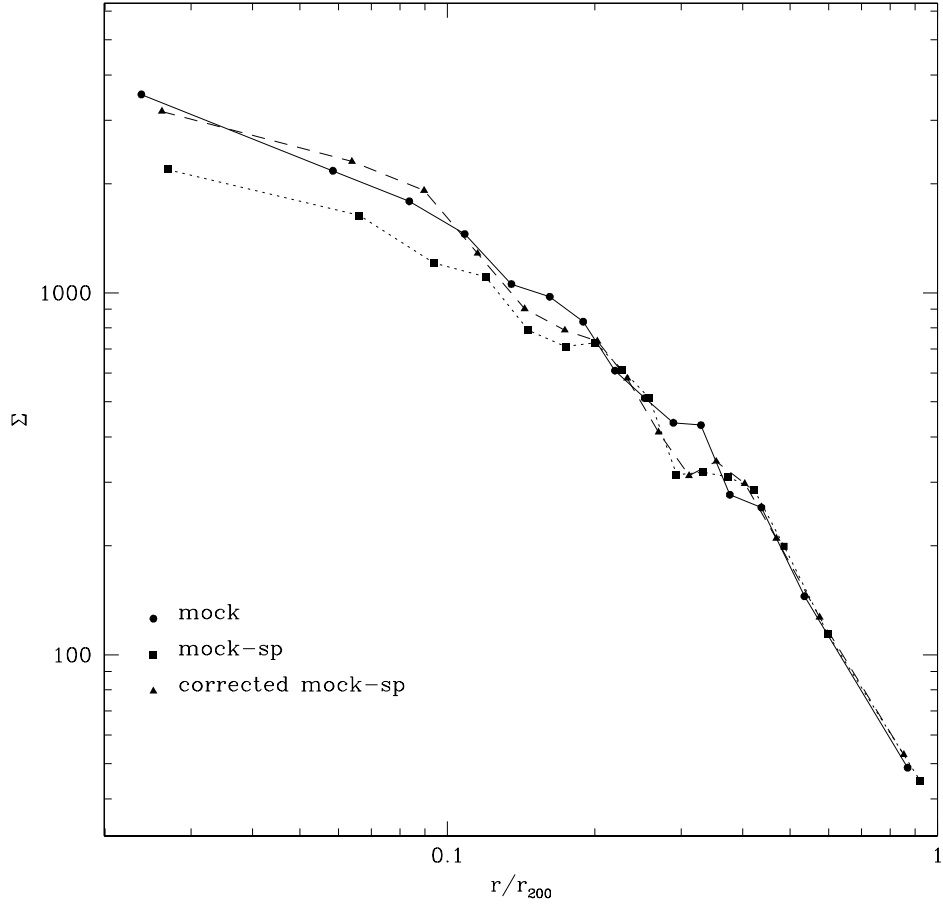}
\caption{
Projected density profiles measured from mock catalogs of the SDSS. Circles show
the profile from a mock catalog which includes all galaxy pairs, 
squares are the profile of a mock catalog without 70 $\%$ of the pairs
(mock-sp), while triangles are the corresponding profile to the mock-sp 
corrected by the missing close pairs.
\label{figpares}
} 
\end{figure}

In the observational process of the SDSS sample, there is a restriction in the 
targeted objects since the fiber centers can not be placed closer than $55''$ 
on a given plate. This limitation produces the missing-pair problem, so that, 
one of the pair components can not be observed. The loss of galaxies 
 was quantified by Strauss et al. (2002), showing that about the $6\%$ 
of the galaxies are not observed owing to the missing-pair problem. 
This percentage 
represents roughly the $70 \%$ of the total number of galaxy pairs, while 
the remaining $30\%$ was measured due to the overlapping of plates in some 
regions.  
In order to analyze the possible effect of this loss of galaxies on the 
resulting projected density profile we work with mock catalogs. 
We construct two SDSS mock catalogs from N-body simulations (section 2.1) 
following a similar prescription as the adopted for the 2dF mocks, but
using the luminosity function computed by Blanton et al. (2003). 
In one of these catalogs (mock-sp) we reproduce the missing-pair problem of the 
SDSS sample. This was achieved selecting the $70\%$ of the galaxy pairs 
and subsequently we remove one component of each pair in this subsample.
We measure the projected density profiles for both mock catalogs following 
the procedure used for the mock-m of the previous section. Figure 
\ref{figpares} shows the comparison among these density profiles. Circles are 
the projected density profile measured in the SDSS mock catalog with the full 
sample of pairs, while the squares correspond to the profile obtained from 
the mock catalog affected by the missing-pair problem. 
This figure shows that this problem produces a significant variation on the 
projected density profile mainly in the inner regions of groups, biasing 
the sample towards profiles with a core.  
We develop a method to correct this effect adding random galaxies to the 
sample of group galaxies. The outline of the method is as follows:
\begin{enumerate}
\item We seek for the $30\%$ of existing pairs in the galaxy catalog for which
both members were observed: 
$N_1$ (number of galaxies in pairs with distances less than $D_{m_{SDSS}}^{\ast}=55''$). 
\item We identify which of the $N_1$ galaxies belong to groups: $N_2$
\item We calculate the percentage of galaxies in pairs that are group members: 
$P_1=N_2/N_1$. Here we have assumed that the probability of measuring both 
members of a galaxy pair does not depend on whether it is in a group or not.
\item Using that $N_1$ is the $30\%$ of the full sample of pairs, we estimate 
the number of galaxies in pairs that belongs to the remaining $70\%$: 
$N_3=7/3 \times N_1$
\item We calculate how many of the $N_3$ galaxies would be found in groups: 
$N_4=N_3 \times P_1$
\item Finally, the number of galaxies to introduce in the sample is $N_5=N_4/2$
 since we already have one of the pair component in the sample.
\item Using the galaxies in pairs that belongs to groups ($N_2$), we measure 
its redshifts and $r/r_{200}$ distributions relative to the center of the
group to which each of the galaxies belong.
\item For the computation of the projected density profiles, we randomly 
add $N_5$ galaxies reproducing the previous distributions.
\end{enumerate}
We apply this procedure to the mock-sp, and measure the corresponding density 
profile, which is also shown in Figure \ref{figpares} (triangles). It can be 
observed that our method is capable to correct the missing-pair effect on 
density profiles. We will use this method in the SDSS group sample in the 
following sections in order to obtain a fair estimate of the projected density 
profiles.

\begin{figure}
\plotone{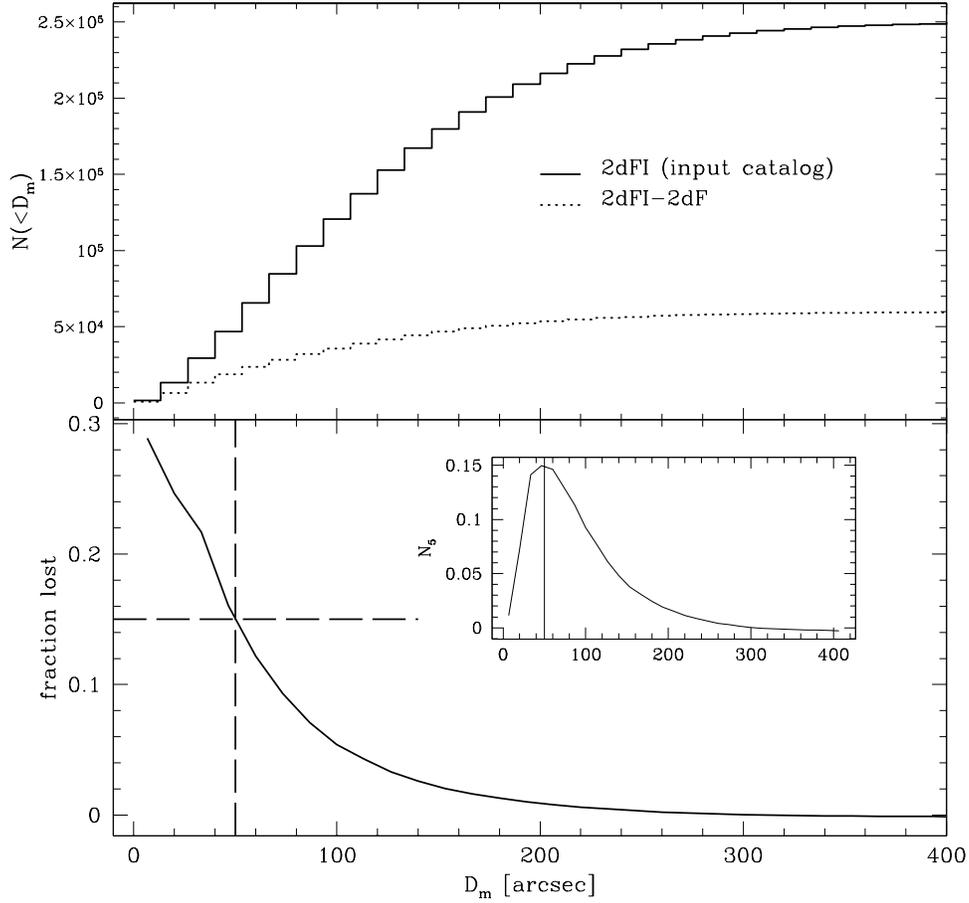}
\caption{
Upper panel: Cumulative galaxy distributions as a function of the angular 
distance to the nearest neighbor. Solid line represents the distribution 
for 2dFGRS input catalog (2dFI) while dotted line 
shows the distribution for 2dFI galaxies that are not included in 
the 2dFGRS. Lower panel: Cumulative fraction of 
lost galaxies in the 2dFGRS as a function of the minimal angular distance. 
Vertical dashed line determines the maximum angular distance $D_{m_{2dF}}^{\ast}$ 
adopted for the missing-pair correction in the 2dFGRS. Inset panel: 
Distribution of missed galaxies in groups as a function of $D_m$ due to the missing-pair problem. 
\label{figpares2}
} 
\end{figure}

For correcting the missing-pair problem in the 2dF sample we apply the 8-step 
procedure described above where the values used for the SDSS must be changed: 
in items 1 and 4 $(30\%)_{SDSS} \rightarrow (85\%)_{2dF}$; 
item 4 $(70\%)_{SDSS} \rightarrow (15\%)_{2dF}$ and in item 1 
$(55'')_{SDSS} \rightarrow (50'')_{2dF}$. 
The rest of this subsection describes how we find the percentage of lost 
galaxies in pairs ($15\%$) and the maximum distance to define the 
missing-pair problem ($50''$) for the 2dF sample.
We first measure two distance distributions: the first is the distribution 
of distances from each galaxy to its closest neighbor ($D_m$) 
among galaxies that belong to the input catalog of the 2dFGRS (Colless et al., 
2003); the second distribution was built from a subsample of the previous one.
This subsample (2dFI-2dF) comprises all the galaxy pairs in the input catalog
that were not completely surveyed by the 2dFGRS. 
Their cumulative distributions are shown in the upper panel of Figure 
\ref{figpares2}. 
The solid line corresponds to the input catalog while the dotted line is the
histogram for the subsample 2dFI-2dF. 
The ratio among these cumulative distributions is the fraction
of lost galaxies in the 2dF until a given angular distance. This loss is due
to two issues in the observational process: the sky coverage of the sample and
the missing-pair problem. 
There is a scale beyond which the ratio of missed pairs has to be constant 
as function of $D_m$, as there is for both input catalog and redshift 
catalog a maximal $D_m$ value. 
Indeed, by definition $D_m$ is the minimal distance to a neighbor, which has 
to reach a maximal value in a finite sample. Therefore beyond $D_m^{max}$ the 
ratio is constant. 
That constant value correspond to the incompleteness due to
the sky coverage and it has to be subtracted from the ratio in order 
to obtain an estimate of the close pair incompleteness. 

The resulting fraction as a function of $D_m$ is shown in the 
lower panel of Figure \ref{figpares2}. 
From this plot we should be able to determine the fraction of lost galaxies
but, firstly, it is necessary to know the angular distance $D_{m_{2dF}}^{\ast}$
so it is representative of the largest number of the galaxies that
were missed due to the missing-pair problem.
In order to determine $D_{m_{2dF}}^{\ast}$ we calculate the number of galaxies
that must be added until a given angular distance $D_m$.
These numbers are calculated for each $D_m$ following the steps 1 to 6, 
previously described for the correction of the missing-pair problem. 
The fraction of lost galaxies involved in this procedure are obtained
from the lower panel of Figure \ref{figpares2}.
Then, the resulting numbers of galaxies that must be added until a given 
angular distance $D_m$ are shown in the inset panel of the 
lower panel of Figure \ref{figpares2}. 
From this distribution we determine $D_{m_{2dF}}^{\ast}=50''$ 
that corresponds to 
the $D_m$ where the distribution is maximum.
This is the optimal way to determine the $D_{m_{2dF}}^{\ast}$ since this value
allows us to compute the appropriate correction to the projected density 
profiles. This was tested constructing a mock catalog with a similar sky 
coverage to that observed in the 2dFGRS and an enhanced pair incompleteness. 
The later characteristic is necessary to obtain a large enough effect so 
it can be measured in the projected density profile.
As in the case of the SDSS, we observe that the missing-pair 
underestimates the amplitude of the density profile in the inner region.
After performing a similar analysis to that shown in Figure 
\ref{figpares2} but using the mock catalog, we observe that introducing the 
maximum number of galaxies (ie., the peak of the distribution in the lower 
inset panel) allow us to recover the true density profile.
This result confirms that the procedure adopted to obtain the value of 
$D_{m_{2dF}}^{\ast}$ is the optimal.
To conclude, we observe that using $D_{m_{2dF}}^{\ast}=50''$ means that the 2dFGRS 
losses approximately $15\%$ of the galaxies (it implies that 2dF has lost
$N_5=87$ pair-members from the sample of galaxies in groups used in this work). 

The missing-pair correction is an important issue to be considered when
working on the SDSS (it looses the $70\%$ of the galaxy pairs). The
density profiles with and without this correction are different mainly in 
the inner regions. On the other hand, the correction applied to the 2dFGRS
($15\%$ of the galaxy pairs) will not introduce a significant change in the
resulting density profile (see Fig. \ref{fig6} and references in the 
next section). 

\subsection{2dF and SDSS projected density profile}
Once we have corrected the samples by the missing-pair problem, 
we construct the composite samples and  split galaxies in 3 spectral types.
For the 2dF sample we use the classification made by Madgwick et al.(2002) 
to spectral types. 
This classification is determined by the shape of the 
$\eta$-distribution:
\begin{itemize}
\item Type 1: $~~~~~~~~~~~\eta < -1.4$,
\item Type 2: $-1.4\leq \eta < ~~1.1$,
\item Type 3+4: $~~~~~~~\eta\ge ~~1.1$.
\end{itemize}
The first type is dominated by elliptical and early-type 
spiral galaxies, getting later toward type 3+4. 

\begin{figure}
\plotone{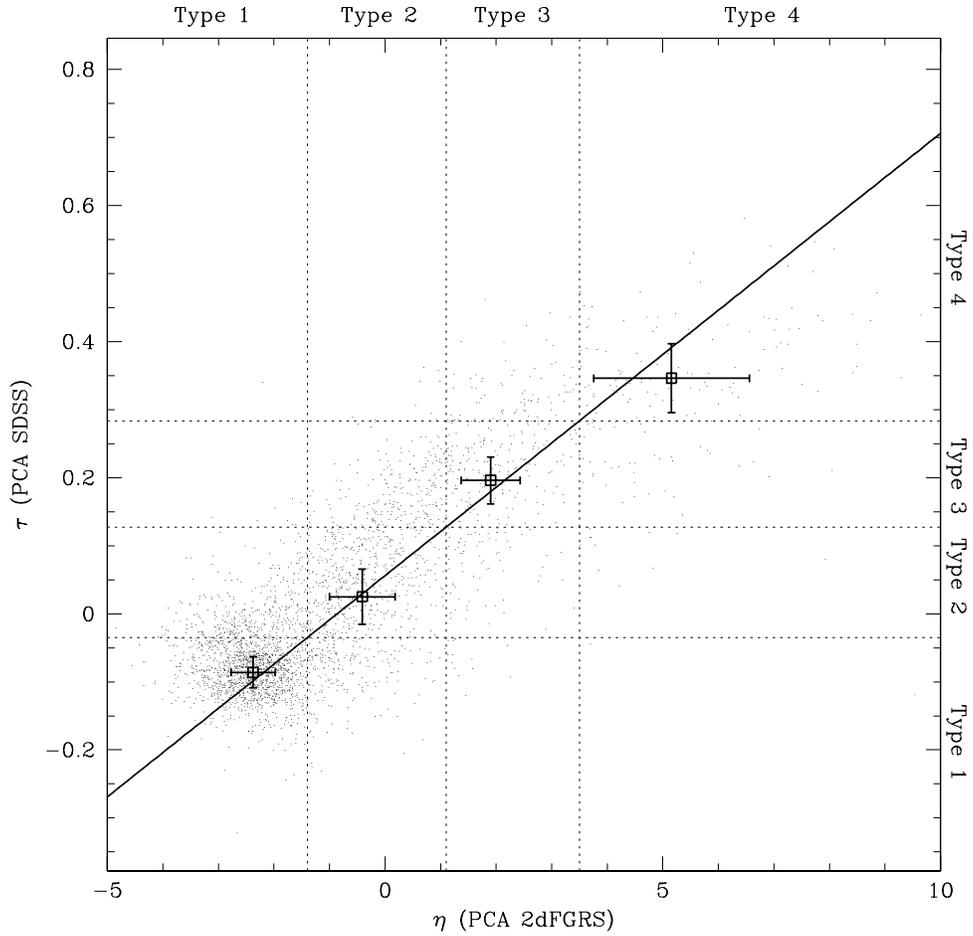}
\caption{
Correlation between the PCA spectral parameter $\eta$ (2dFGRS) and $\tau$ 
(SDSS). The straight line is the linear fit to the data points. The open squares
are the median $\eta$ and $\tau$ values per spectral type, 
whereas the error bars are the semi-interquartile ranges. 
\label{recta}
} 
\end{figure}

In an attempt to obtain a similar spectral type classification for the 
SDSS group sample, 
we seek for a correlation between the spectral parameters 
$\eta$ of the 2dF and $\tau$ of the SDSS comparing $3300$ galaxies that both
have in common. We find a roughly linear behavior among both parameters 
(Figure \ref{recta}). The fit obtained for this relation is: 
\begin{equation}
\tau=(0.065\pm0.002)\ \times \eta+(0.056\pm0.003)
\end{equation} 

Using this relation we divide the SDSS sample into 3  
spectral types according to the classification made for the 2dF. 
Hence the resulting classification for the SDSS sample is: 
\begin{itemize}
\item Type 1: $~~~~~~~~~~~~\tau < -0.035$,
\item Type 2: $-0.035\leq \tau < ~~0.128$,
\item Type 3+4: $~~~~~~~~\tau\ge ~~0.128$.
\end{itemize}
We measure the projected density profiles for the composite samples
(the whole sample and the samples selected per spectral type). The 
procedures to introduce weights in the estimation are the same as the used in
mock-c-$m_v$ for the 2dF and using the equation \ref{w(z)} for the SDSS.  
For the 2dF sample we adopt the luminosity functions per spectral type 
of galaxies in groups given by Mart\'{\i}nez et al. (2002) and $k+e$ 
corrections given by Norberg et al. 
(2002). For the SDSS we use the luminosity functions per spectral type of 
galaxies in groups 
estimated following the same procedure as described by Mart\'{\i}nez et al.
(2002). The $k+e$ corrections as a function of redshift were estimated 
following a method similar to that described by Norberg et al. (2002), 
using the code of stellar population synthesis developed by Bruzual \& 
Charlot (1993). 

\begin{figure}
\plotone{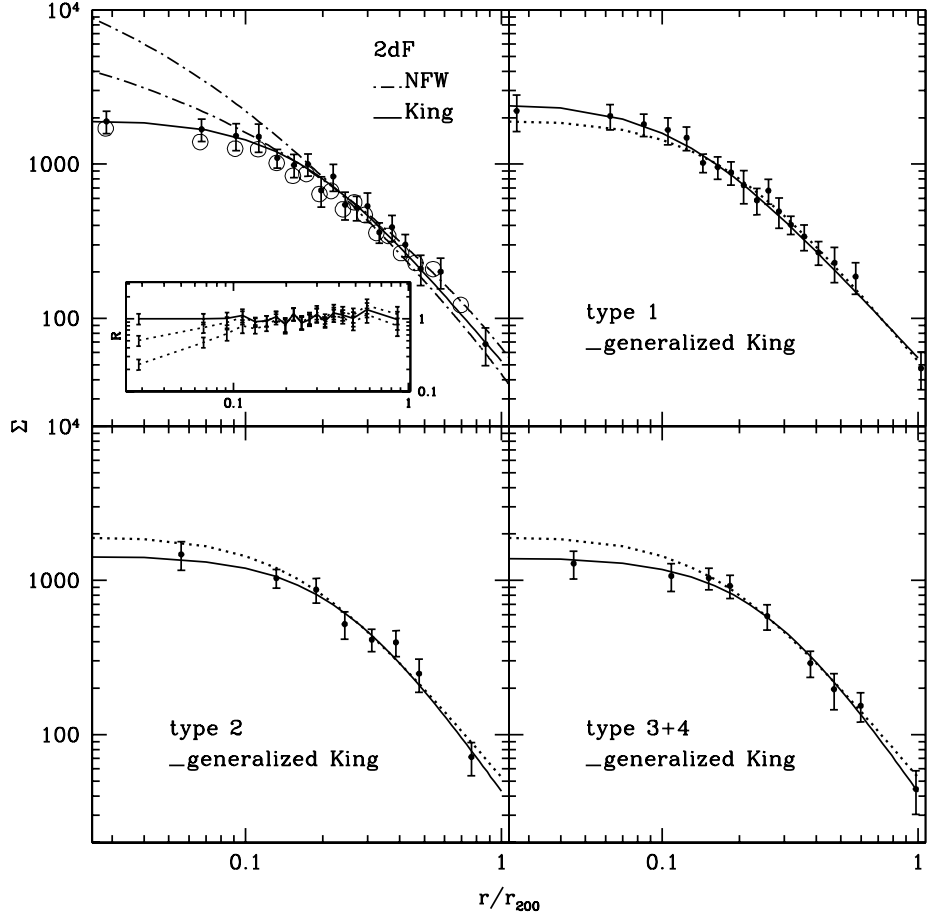}
\caption{
2dF projected galaxy density profiles as a function of a normalized 
group-centric distance. Left upper panel shows the projected density profile (filled circles) measured for the whole composite sample. Open circles are the profile measured without introducing the missing-pair correction. 
In this panel, dot-dashed lines are
the projected NFW profiles that expands the range of masses of physical 
relevance as the shown in Figure \ref{fig1}.
 Projected density profiles for 
spectral types $1$, $2$, $3+4$ are shown in right upper, left lower and right 
lower panels respectively. The solid lines in each panel are King and 
generalized King fits 
with best-fitting parameters quoted in Table \ref{2Dtab}.
We also plot with dotted line in the panels per spectral type the profile 
fitted for the complete composite sample. 
The inset box in the left upper panel shows the ratio between the observational 
projected density profile  and the best-fit King (solid line) and the two 
reference NFW (dotted lines) profiles. 
\label{fig6}
} 
\end{figure}

\begin{figure}
\plotone{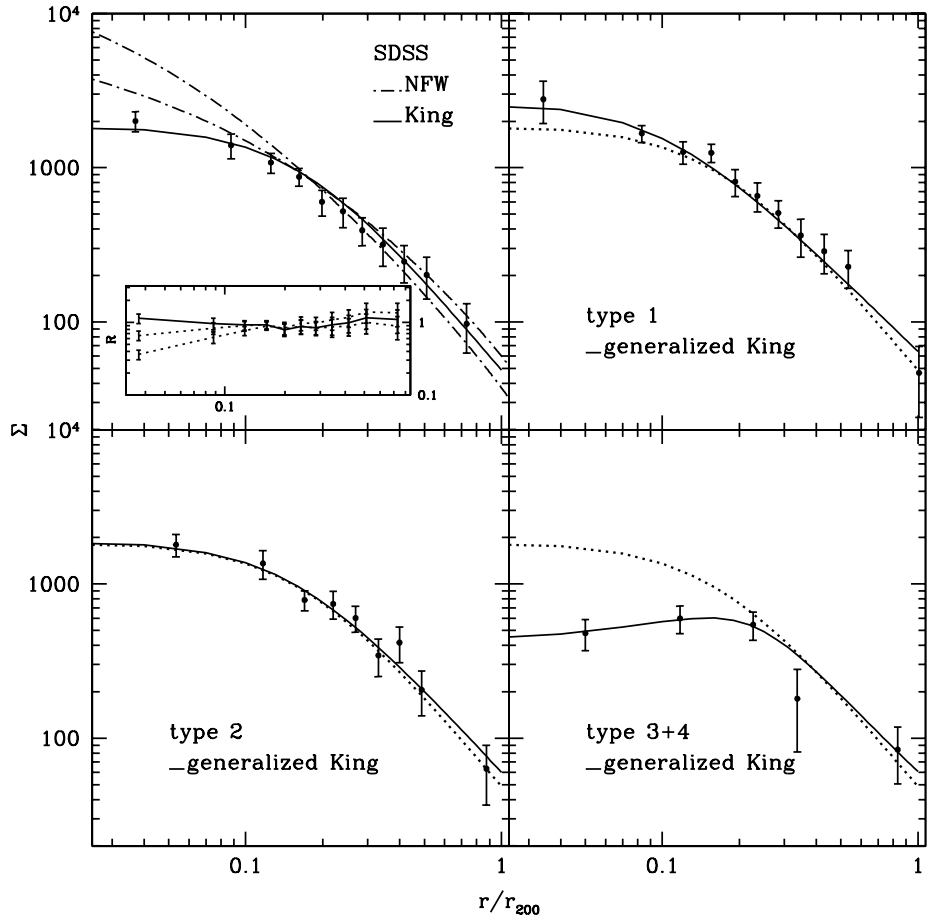}
\caption{
Same as Figure \ref{fig6} but measured for the SDSS composite sample.
\label{fig7}
} 
\end{figure}

Figures \ref{fig6} and \ref{fig7} show projected density 
profiles for the 2dF and the SDSS composite samples, respectively. 
The left upper panels show the profiles for the whole sample (points). The open 
circles are the projected density profile measured without correcting by 
the missing-pair problem. As we mentioned in the previous section, the effect 
is not significant for this sample.  
Error bars in these panels were calculated computing the mean dispersion 
obtained using sets of 10 mock catalogs.
The remaining three panels correspond to projected density profiles per 
spectral types (points) where error bars are computed performing a bootstrap
resampling of the data.

In the left upper panels of both Figures we confront the measured projected 
density profiles for the complete samples against two analytical  
NFW projected density profiles (dot-dashed lines). 
Any profile corresponding to groups with masses between $10^{11} M_\odot$ 
and $10^{15} M_\odot$ should lie in the region determined by these two 
NFW profiles.
From the comparison, we can notice that the dark matter NFW profile 
does not show the same behavior as the obtained from galaxy samples in the 
inner regions of the galaxy groups.
We also compare our results with the analytical prediction for galaxy 
density profiles given by King (1962).
We fit the data points using a generalized King profile given by the formula:
\begin{equation}
\Sigma_m (r)\propto \frac{1}{(1+(c(\frac{r}{r_{200}}-x_0))^2)^\beta}
\label{king}
\end{equation}
where $c$ is the concentration parameter defined as $r_{200}/r_{core}$, $\beta$
is the slope in the outer region and $x_0$ is the radius where the profile 
reaches its maximum value. These parameters are determined using the
Levenberg-Marquardt method (Press et al., 1986). This method takes into account 
data errors and applies a minimum non-linear least squares procedure. 
The number of parameters to fit is chosen to be as low as possible.  
Notice that the King (1962) projected profile is obtained by setting 
$\beta=1$ and $x_0=0$. Using a mean value of $r_{200}$, and a range for 
$r_{core}$ from $100$ to $300$ $kpc \ h^{-1}$ then the allowed values of 
$c$ parameter are in the range $4-13$. 
The best-fitting parameters obtained for the full 
sample and for each spectral type are listed in Table \ref{2Dtab}.
These fits are shown with solid lines in Figures \ref{fig6} and \ref{fig7}. 
The dotted lines show the King-fit obtained for the whole sample.\\ 
In order to quantify the goodness of the fits we compute
the probability $Q$ which is a function of the $\chi^2$ and the degree of
freedom of the distribution $\nu$ (Press et al., 1986).
The chi-square probability $Q(\chi^2,\nu)$ is an incomplete gamma function and
gives the probability that the chi-square should exceed a particular value 
$\chi^2$ by chance. For a fit with $M$ free parameters (${\bf{a}}$) 
the $\chi^2$ is calculated using
\begin{equation}
\chi^2=\sum^{N_{bin}}_{i=1}\left(\frac{\Sigma_i-\Sigma_{m}(\frac{r}{r_{200}};\bf{a})}{\sigma_i}\right)^2; \ \ \ {\bf{a}}=(a_1,a_2,...,a_M)
\end{equation}
while the number of degrees of freedom of the distribution is computed as
$\nu=N_{bin}-M$.
Using the $Q(\chi^2,\nu)$ value, the goodness of a fit is quantified in
the following way: a value of $Q > 0.1$ says that the fit is a very good 
reproduction of the data distribution; if the value is in the range 
$0.001<Q<0.1$ then the fit is acceptable and finally, if $Q<0.001$ the model 
poorly fit the data. It should be remarked that this kind of test is also
valid even when the models are not strictly linear in the $a's$ coefficients.
In the last column of Table \ref{2Dtab} we quoted the
$Q$ values of all fits in both samples. From these values we conclude that 
almost all the fits obtained are a very good approximation to the measured 
projected density profiles and only one of them is in the range of acceptable. 
As an important result we can observe, in left upper panel of Figures \ref{fig6}
and \ref{fig7}, that the King profile is a good descriptor of the observational 
data in the whole range of $r/r_{200}$ and the $c$ values are within the 
allowed range. This result is consistent with the obtained
by Adami et al. (1998) who found that the King profile provides a better
fit to the galaxy density profile than the NFW profile. 
This result can be more clearly observed in the inset box in the upper left 
panel of these figures. In these panels we show the ratio between the 
observational projected density profile and the King profile (solid line). In 
dotted lines are also shown the ratio between the observational density 
profile and the NFW predictions. These figures show that NFW predictions differ 
from unity in the inner regions whereas King fits are roughly constant in 
the whole range.

\begin{figure}
\plotone{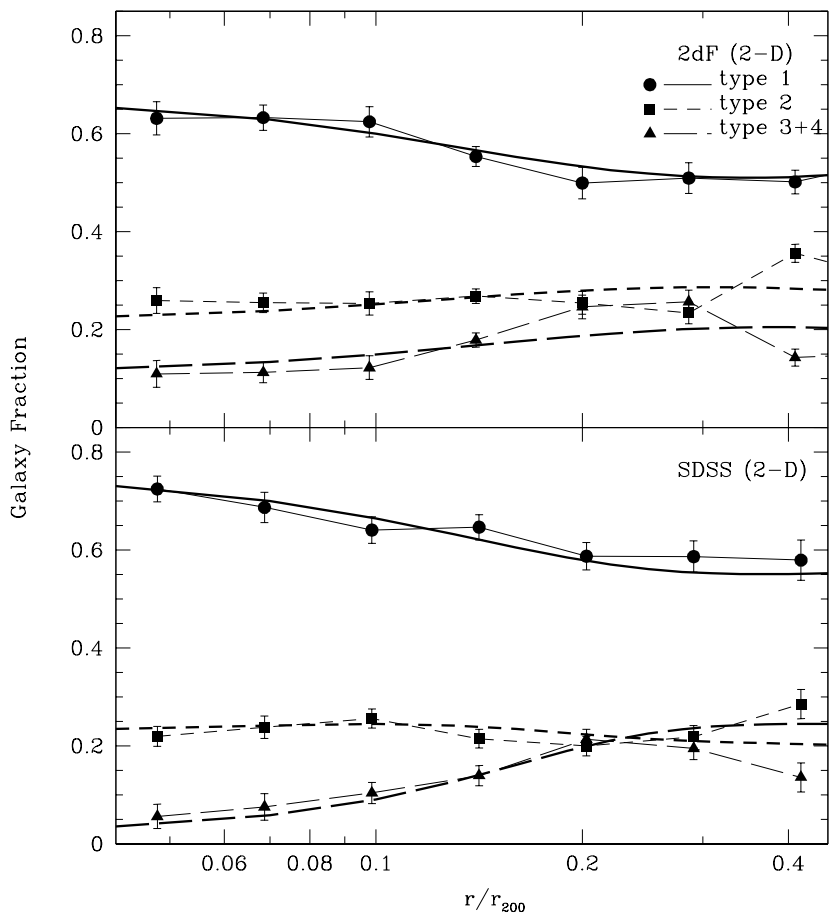}
\caption{
2-D fractions of galaxies of different spectral types as a function of the 
normalized group-centric distance for the 2dF (upper panel) and the SDSS (lower panel). The key for the 
different spectral types is included in the figure. Error bars are computed 
from error propagation. Thick lines correspond to the galaxy fractions
computed from the generalized King profiles fitted to the projected density
profiles.
\label{fig8}
} 
\end{figure}

Based on the projected density profiles for each spectral type, we calculate 
the relative fraction of galaxies with different spectral types as a function 
of the projected group-centric distance. This fraction is computed as the ratio 
between the projected density profile for a particular spectral type and 
the total projected density profile. In order to measure these ratios, we 
have rebinned the data, using a linear interpolation.
Figure \ref{fig8} shows the galaxy fractions for the 2dF composite sample (upper panel) and  for the SDSS composite sample (lower panel). 
Using error propagation method, the error in the relative 
fraction of the type $j$ for each bin is defined by the following equation:
\begin{equation}
\sigma_{frac_j}=\sqrt{(\sigma_{\Sigma_j}/\Sigma_T)^2+(\sigma_T \Sigma_j/\Sigma_T^2)^2}
\label{error}
\end{equation}
where $\Sigma$ represents the projected density profile, and $\sigma_\Sigma$ 
represents the error in these profiles.   
Thick lines in these Figures 
correspond to galaxy fractions calculated using the fits (Table \ref{2Dtab}) 
obtained for the projected density profiles per spectral type.
The results are similar to that found by other authors when considering 
spectral types (Dom\'{\i}nguez et al. 2002) or morphological 
classifications (Whitmore \& Gilmore 1991, WGJ93).
We observe in both, the 2dF and the SDSS group sample, that for 
small $r/r_{200}$ radii the fraction of early type galaxies (Type 1)
rises rapidly whilst the fraction of late type galaxies tends 
to be more important in the outerskirt of the galaxy systems.
By comparing these panels we observe that the behavior of each type 
for both samples is quite similar and the main difference is only in the 
amplitudes. This difference is expectable because the samples are selected in 
different band magnitudes: $r-$band for the SDSS and $b_j-$band for the 2dFGRS.
Therefore the percentage of low star forming galaxies (type 1) is going to 
be higher in the SDSS than in the 2dF whilst the opposite is found for the 
star forming galaxies (type 3+4).

\section{3-D galaxy fractions}
\begin{figure}
\plotone{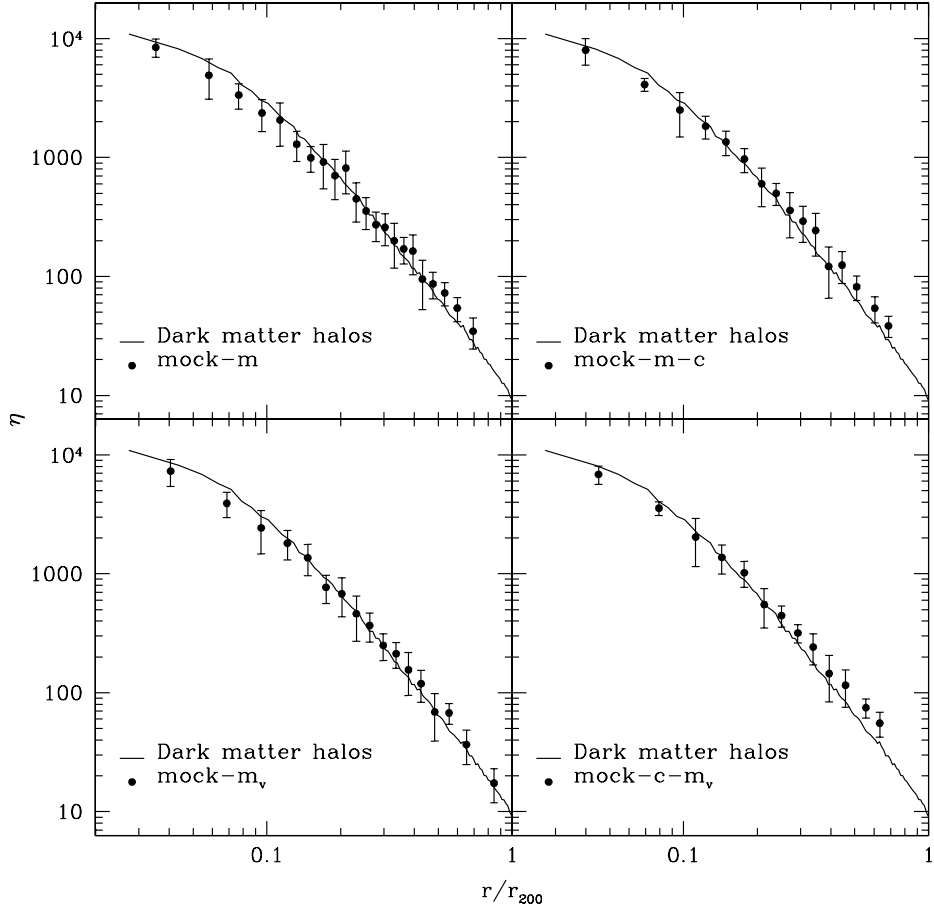}
\caption{
Average deprojected galaxy density profiles obtained from the projected galaxy 
density profiles of mock catalogs showed in Figure \ref{fig4}. 
The points in the left upper panel show the average value of the deprojection 
of ten 2-D profiles obtained from mock-m catalogs. Error bars are the 
associated $1-\sigma$ dispersion.
Right upper, left 
lower and right lower panels show the same profiles as in the left upper panel 
but obtained from mock-m-c, mock-$m_v$ and mock-c-$m_v$ respectively. 
Solid line in each panel corresponds to the deprojected density
profile obtained from the projected density profile 
computed for groups identified in the dark matter N-body simulations.
\label{fig5}
} 
\end{figure}

\begin{figure}
\plotone{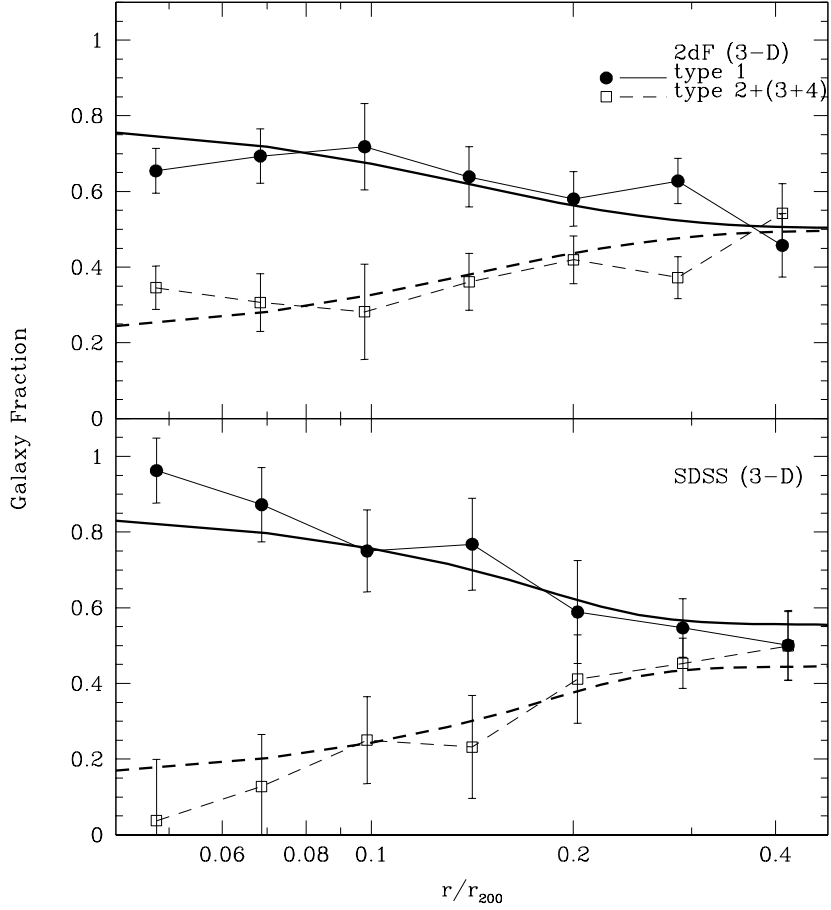}
\caption{
3-D fraction of galaxies per spectral type as a function of the normalized 
group-centric distance for the 2dF (Upper panel) and for the SDSS (Lower panel).
The key for the different spectral types is included in the figure. Error 
bars are computed from error propagation. Thick lines correspond to the 
galaxy fractions computed from the generalized King profiles (see text).
\label{fig12}
} 
\end{figure}

One of the aims of this work is to derive information about the 3-D galaxy 
distributions from observational data. To achieve this aim we use the 
deprojection method described in section 2 to obtain the 3-D galaxy 
density profiles from the projected ones. 
The deprojection method produces very good estimates when we are
dealing with projected profiles without bin to bin fluctuations, 
as it was the case in section 2 for N-body simulations. 
Nevertheless, if the profiles show bin to bin fluctuations,
this method tends to enhance the noise in the resulting profiles, 
from outer to inner radii of galaxy systems. 
For instance, Figure \ref{fig5} shows the averaged deprojected density 
profiles corresponding to the deprojection of the 2-D profiles of ten 
mock catalogs of each kind (the averages of these 2-D profiles are shown 
in Figure \ref{fig4}).
It can be seen that 3-D profiles present an important dispersion. 
Therefore, it must be expected that the intrinsic noise observed for the 
projected density profiles in the 2dF and SDSS catalogs will be 
amplified by the deprojection method.
We carry out the deprojection of all the observed 2-D profiles measured in the 
previous section, obtaining noisy 3-D profiles. In order to estimate errors to 
these profiles, we perform bootstrap resampling of the 2-D data and then we 
deproject each bootstrap density profile. By calculating the 1-$\sigma$ 
dispersion of the 3-D bootstrap profiles, the errors for the 3-D profile of 
the data are obtained.
From the deprojected profiles we calculate the 3-D galaxy fraction per spectral 
 type. These fractions can be seen in Figure \ref{fig12} for the 2dF 
(upper panel) and SDSS (lower panel) samples, where the error bars are 
calculated by error propagation as in the 2-D case (equation \ref{error}). 
We decide to join the types 2 and (3+4) because the resulting noise of 
the deprojection does not allow to observe differences among them. 
It should be noticed that the 
importance of this result resides in the fact that these 3-D fractions are 
obtained by directly inverting the 2-D profiles, it means, without assuming 
a particular shape for the density profile.

Finally, we also show in Figure \ref{fig12} the 3-D fractions per 
spectral type calculated analytically (thick lines). 
The analytical profiles that are needed for the computation of these fractions 
come from integrating the analytical fits of the 2-D density profiles assuming 
spherical symmetry. The result of the integration shows that the 3-D 
profiles keep the same functional form 
that the adopted for the 2-D case (generalized King), with values of 
$c_{3D}=c_{2D}$ and $\beta_{3D}=\beta_{2D}+0.5$. In this Figure it can 
be observed that the analytical curves present the same behavior that those 
obtained from the direct deprojection, indicating that the generalized King 
density profiles are also a good fit for the observational result 
in the 3-D case.

\section{Summary and Conclusions}
Using the two largest galaxy redshift surveys presently available, the final 
release of the 2dFGRS and the first release of the SDSS, we carry out an 
analysis of the galaxy populations and their distribution in massive groups of 
galaxies. Group identification on these surveys is made using an 
algorithm similar to that developed by Huchra \& Geller (1982). 
Particularly, for the 2dFGRS sample, we introduce modifications, 
in order to take into account the non-uniform sky coverage of this survey 
(Merch\'an \& Zandivarez, 2002). From a careful study of groups identified 
in mock catalogs we realized that this method could 
produce false identifications in redshift space, producing an artificial 
enhancement of the group sizes or merging small systems in larges artificial 
groups. To solve these problems we identify groups upon 
the previous group sample, varying the density contrast until the redshift 
space identification is capable to reproduce the identification 
obtained in real space using $\delta \rho / \bar{\rho} =80$. The new 
density contrast found in mock catalogs is used to perform a second 
identification in the 2dF and SDSS group samples. The group centers are
estimated using an iterative method, which is capable to locate the group
centers upon the overdensity peaks. We also correct the group samples
for the missing-pair problem.

Once we have reliable group samples, we proceed with the analysis of the 
galaxy distribution in galaxy groups. This analysis comprise the study 
of density profiles for high mass groups. These profiles are derived using 
composite samples which are a combination of all groups in each catalogs. 
The normalization scale used to conform the composite sample is $r_{200}$,
the radius at which the interior density is 200 times the mean density of the 
universe, and the mass normalization is $M_{200}$. The results found in this 
work do not depend on the normalization scale.   
Since the group samples used in this work are magnitude limited, our estimator 
of the projected galaxy density profiles is developed introducing 
weights in the galaxy counts which take into account a fixed or variable 
apparent magnitude limit and variations in the redshift completeness. 
The way of introducing these corrections was tested in mock catalogs, 
obtaining a good agreement with the density profiles derived from DM groups 
identified in the N-body simulations. 
The galaxy projected density profiles obtained for the 2dF and SDSS show a 
similar 
behavior. From the comparison of our results with the analytical projected NFW, 
we observe that the last fails to describe the behavior of the observational 
results in the inner region of groups.
This seems to indicate that the dark matter profile (NFW) is not appropriated 
to describe the density profile traced by the luminous matter in the
very core of galaxy groups. 
We observe that the King profile is a better fit for the observational data
in the whole range of group-centric distances where the profiles can be 
measured, in agreement with the results of Adami et al. (1998). 
This result is also found when the normalization scale is modified, for 
instance, using any intrinsic projected group size as a normalization scale.
We tested this point using the virial radii and the rms projected 
physical separation of galaxies from the group center.
The use of these new normalization scales ($r_{new}$) 
produces the same $\beta_{new}$ parameters that we obtain using $r_{200}$ as 
a normalization scale. The $c_{new}$ parameters  are just a re-scaling of the previous and they are given by:
$c_{new}=c_{200} \times \frac{r_{new}}{r_{200}}$.

We obtain that the resulting density profiles are reliable for 
$r/r_{200} > 0.03$, it was calculated taking into account the minimum distance between pairs of galaxies 
or the mean galaxy size, both at the mean distance of the galaxy systems. 
This constrain does not change our results since the fits are made from
$r/r_{200}=0.03$ to higher values. Then, we can claim that the presence
of a core in the projected density profiles is an intrinsic property of
these galaxy systems and not a result of a biased measurement.

We also measure the projected density profiles per spectral type in both 
samples. In order to obtain analytical functions to describe the observational 
results, we adopt a generalized King profile to fit them.
These results could be used as a tool to constrain semi-analytical models. 
The general galaxy density profile and the dependence of the density 
profiles on the spectral types must be correctly reproduced by the models.

Based on the available spectral type information, we compute the 
galaxy fractions per spectral type as a function of the normalized 
group-centric distance $r/r_{200}$.
Our results are in good agreement with the previously obtained by other 
authors (Whitmore \& Gilmore, 1991, WGJ93, Dom\'{\i}nguez et al., 2002): 
the fraction of early type galaxies decreases when $r/r_{200}$ increases 
whereas the opposite behavior is observed for the fraction of later types.

Using the obtained 2-D density profiles, we calculate the 3-D galaxy density 
profiles from their projected counterpart using a deprojection method similar 
to the one developed by Allen \& Fabian (1997).
The 3-D galaxy fractions are computed from the deprojected density profiles per
spectral type. 
By comparing the 2-D and 3-D galaxy fractions it can be seen that the MS effect 
is more pronounced when 3-D fractions are analysed. Finally, 
the analytical 3-D fractions are calculated from the fits obtained for the 
2-D density profiles. It is found a good agreement with the 3-D fractions 
directly calculated without assuming an analytical 2-D density profile.

\acknowledgments

Special thanks to the anonymous referee, for helping us to improve the original version of this work.
We thanks to Juli\'an Mart\'{\i}nez for help us in the 
calculation of the $k+e$ correction for the SDSS and Cinthia Ragone for 
reading the manuscript. 
We also thank to Peder Norberg and Shaun Cole for kindly providing the
software describing the mask of the 2dFGRS and to the 2dFGRS and SDSS Team
for having made available the actual data sets of the sample.
This work has been partially supported by Consejo de Investigaciones 
Cient\'{\i}ficas y T\'ecnicas de la Rep\'ublica Argentina (CONICET), the
Agencia Nacional de Promoci\'on Cient\'{\i}fica, the
Secretar\'{\i}a de Ciencia y T\'ecnica de la Universidad Nacional de C\'ordoba
(SeCyT), the Agencia C\'ordoba Ciencia and Fundaci\'on Antorchas, Argentina.


\clearpage

\clearpage

\clearpage

\clearpage

\clearpage

\clearpage

\clearpage

\clearpage

\clearpage

\clearpage

\clearpage


\clearpage

\clearpage

\begin{deluxetable}{ccccccccc} 
\tablecolumns{9} 
\tablewidth{0pc} 
\tablecaption{Median group properties and distribution widths for the group samples.
\label{proptab}}
\vskip 1cm
\tablehead{ 
\colhead{}    &  \multicolumn{4}{c}{2dFGRS} &
\multicolumn{4}{c}{SDSS} \\ 
\cline{2-9} 
\colhead{} & \colhead{$z$}   & \colhead{$\sigma$}    & \colhead{$M_{vir}$} & \colhead{$R_{vir}$} \vline &  \colhead{$z$}   & \colhead{$\sigma$}    & \colhead{$M_{vir}$} & \colhead{$R_{vir}$} \\   
\colhead{} & \colhead{}   & \colhead{$[km \ s^{-1}]$} & \colhead{$[M_\odot \ h^{-1}]$} & \colhead{$[Mpc \ h^{-1}]$} \vline & \colhead{} & \colhead{$[km \ s^{-1}]$} & \colhead{$[M_\odot \ h^{-1}]$} & \colhead{$[Mpc \ h^{-1}]$}}  
\startdata 
median & $0.09$ & 442 & $8.7\times 10^{13}$ & 0.9 &$ 0.08$ & 427 & $9.3\times 10^{13}$ & 0.9 \\
width  & $0.03$ & $133$ & $1.2\times 10^{14}$ & 0.3 &$ 0.03$ & 159& $1.6\times 10^{14}$ & 0.4 \\
\tableline

\enddata
\end{deluxetable}

\begin{table}
\begin{center}
\caption{Best-fitting parameters for the projected galaxy density profiles. Parameters without error were fixed in the fitting process. The last three columns 
show the degrees of freedom, the chi-square value of the fits and the $Q$ 
probability.
\label{2Dtab}}
\vskip 1cm
\begin {tabular}{clccccccc}
\tableline 
\tableline 
  Catalog & sample &Ngal& $c$ & $\beta$ & $x_0$ & $\nu$&$\chi^2$ & $Q$ \\
\tableline 
        &  total    &2277& $6.0\pm0.2$ & $1.0$      & $0.0$ &17& 13.73 & 0.68 \\
 2dF    &  type 1   &1710& $8.3\pm0.3$ & $0.90\pm 0.05$ & $0.0$ &16& 9.55 & 0.89 \\
(132)   &  type 2   &339& $4.2\pm0.5$ & $1.2\pm0.1$    & $0.0$ &6& 7.13  & 0.31 \\
        &  type 3+4 &228& $4.0\pm0.4$ & $1.2\pm0.1$    & $0.0$ &7& 4.46  & 0.72 \\
\tableline
      &  total    &2031& $6.1\pm0.3$ & $1.0$   & $0.0$        &10 & 10.50 & 0.40 \\
 SDSS &  type 1   &1543& $9.8\pm0.2$ & $0.8\pm0.1$ & $0.0$  &9& 10.54 & 0.31 \\
 (86) &  type 2   &364& $6.6\pm0.5$ & $0.9\pm0.1$ & $0.0$   &7& 8.54  & 0.29 \\
      &  type 3+4 &124& $5.8\pm0.8$ & $0.7\pm0.2$ & $0.15\pm0.06$&2&6.02& 0.05 \\
\tableline
\end{tabular}
\end{center}
\end{table}

\end{document}